\documentclass[letterpaper]{article} % DO NOT CHANGE THIS
\usepackage{aaai25}  % DO NOT CHANGE THIS
\usepackage{times}  % DO NOT CHANGE THIS
\usepackage{helvet}  % DO NOT CHANGE THIS
\usepackage{courier}  % DO NOT CHANGE THIS
\usepackage[hyphens]{url}  % DO NOT CHANGE THIS
\usepackage{graphicx} % DO NOT CHANGE THIS
\usepackage{natbib}  % DO NOT CHANGE THIS AND DO NOT ADD ANY OPTIONS TO IT
\usepackage{caption} % DO NOT CHANGE THIS AND DO NOT ADD ANY OPTIONS TO IT
\usepackage{subcaption}
\usepackage{multirow} % For the large descriptive stats table - remove if we don't use that table
\usepackage{comment}
\usepackage{tabularx}

\usepackage{booktabs} %for table layouts

\title{\textit{Highlight All the Phrases}: Enhancing LLM Transparency through Visual Factuality Indicators}

\author {
    Hyo Jin Do\textsuperscript{\rm 1}\equalcontrib,
    Rachel Ostrand\textsuperscript{\rm 2}\equalcontrib,
    Werner Geyer\textsuperscript{\rm 1}, \\
    Keerthiram Murugesan\textsuperscript{\rm 2},
    Dennis Wei\textsuperscript{\rm 3},
    Justin Weisz\textsuperscript{\rm 2}
    \\
}
\affiliations {
    \textsuperscript{\rm 1}IBM Research, Cambridge, MA, USA,\\
    \textsuperscript{\rm 2}IBM Research, Yorktown Heights, NY, USA,\\
    \textsuperscript{\rm 3}IBM Research, San Jose, CA, USA\\
    hjdo@ibm.com, 
    rachel.ostrand@ibm.com,
    werner.geyer@us.ibm.com, \\
    keerthiram.murugesan@ibm.com, 
    dwei@us.ibm.com,
    jweisz@us.ibm.com

}

\begin{document}

\maketitle

\begin{abstract}
Large language models (LLMs) are susceptible to generating inaccurate or false information, often referred to as ``hallucinations'' or ``confabulations.'' While several technical advancements have been made to detect hallucinated content by assessing the factuality of the model's responses, there is still limited research on how to effectively communicate this information to users. To address this gap, we conducted two scenario-based experiments with a total of 208 participants to systematically compare the effects of various design strategies for communicating factuality scores by assessing participants' ratings of trust, ease in validating response accuracy, and preference. Our findings reveal that participants preferred and trusted a design in which all phrases within a response were color-coded based on factuality scores. Participants also found it easier to validate accuracy of the response in this style compared to a baseline with no style applied. Our study offers practical design guidelines for LLM application developers and designers, aimed at calibrating user trust, aligning with user preferences, and enhancing users' ability to scrutinize LLM outputs.
\end{abstract}

% Uncomment the following to link to your code, datasets, an extended version or similar.
%
% \begin{links}
%     \link{Code}{https://aaai.org/example/code}
%     \link{Datasets}{https://aaai.org/example/datasets}
%     \link{Extended version}{https://aaai.org/example/extended-version}
% \end{links}

\captionsetup[subfigure]{subrefformat=simple,labelformat=simple}
\renewcommand\thesubfigure{(\alph{subfigure})}

\newcommand{\centeredtxt}[1]{
    \begin{tabular}{l}
        \parbox{2cm}{\vspace{-20pt} \centering #1}
    \end{tabular}
}

% %%
% %% The code below is generated by the tool at http://dl.acm.org/ccs.cfm.
% %% Please copy and paste the code instead of the example below.
% %%
% \begin{CCSXML}
% <ccs2012>
%    <concept>
%     <concept_id>10003120.10003121.10003122.10003334</concept_id>
%        <concept_desc>Human-centered computing~User studies</concept_desc>
%        <concept_significance>500</concept_significance>
%        </concept>
%    <concept>
%        <concept_id>10003120.10003123.10011759</concept_id>
%        <concept_desc>Human-centered computing~Empirical studies in interaction design</concept_desc>
%        <concept_significance>500</concept_significance>
%        </concept>
%    <concept>
%        <concept_id>10003120.10003121.10003124.10010870</concept_id>
%        <concept_desc>Human-centered computing~Natural language interfaces</concept_desc>
%        <concept_significance>500</concept_significance>
%        </concept>
%  </ccs2012>
% \end{CCSXML}

% \ccsdesc[500]{Human-centered computing~User studies}
% \ccsdesc[500]{Human-centered computing~Empirical studies in interaction design}
% \ccsdesc[500]{Human-centered computing~Natural language interfaces}

%%
%% Keywords. The author(s) should pick words that accurately describe
%% the work being presented. Separate the keywords with commas.
%\keywords{Large Language Models, Hallucinations, Factuality, Source Attribution}

%% A "teaser" image appears between the author and affiliation
%% information and the body of the document, and typically spans the
%% page.

\section{Introduction}

Large language models (LLMs) can generate factually incorrect or fabricated information that appears plausible and is presented with confidence -- a phenomenon widely known as ``hallucination'' ~\cite{ji2023survey}. This behavior is also described as ``confabulation'' ~\cite{smith2023hallucination}, or more bluntly, ``bullshit'' ~\cite{hicks2024chatgpt}\footnote{In this paper, we use the term ``hallucination'' to encompass all these concepts; while “confabulation” may be more precise, “hallucination” is more broadly recognized.}. The presence of these hallucinations in LLM outputs, coupled with difficulty in detecting them and users' tendency to over-trust LLMs~\cite{bo2024rely,kim2024m}, has led to several high-profile incidents. For example, lawyers have been reprimanded by judges for referencing hallucinated case law~\cite{Sloan_2023}, new products have been rapidly shelved due to hallucinated scientific references~\cite{ryan22meta}, news outlets have had to issue corrections to articles written with AI assistance~\cite{sato23cnet}, and company share prices have dropped after a hallucination caused a blunder during a new product demo~\cite{Google_Webb}. In response, some communities have prohibited the use of LLM-generated content to safeguard against the inclusion of hallucinated information, such as StackOverflow, an online Q\&A forum \cite{Stackoverflow}.

Researchers are actively exploring ways to mitigate hallucinations by improving datasets and employing techniques such as reinforcement learning with human feedback~\cite{ouyang2022training, ji2023survey} and retrieval-augmented generation~\cite{lewis2020retrieval, cai2022recent}. However, technical advancements alone cannot completely resolve the issue; ultimately, it falls upon end-users to carefully evaluate LLM outputs and be accountable for their use.

Presenting \textit{factuality scores}, which indicate the extent to which a model's response is truthful to a source document ~\cite{kryscinski2020evaluating,laban2022summac,maynez2020faithfulness,zhou2023synthetic, chern2023factool,min2023factscore},
presents a promising human-centered solution to help users in evaluating LLM outputs. (Although definitions of ``factuality'' vary slightly in the field, we use it here to refer to truthfulness with respect to a source document which is considered factual information.)
Nevertheless, the best way to communicate factuality information to users remains unclear. 
As a first step, \citet{leiser2023chatgpt} conducted participatory workshops where participants brainstormed design features to help identify hallucinations in LLM outputs. They found that participants desired either numerical factuality indicators (e.g., percentage) or ordinal factuality indicators (e.g., high, medium, low) with visual aids, such as color-coded underlines to differentiate between factual and fictional arguments.
However, no studies have systematically compared the effectiveness of different strategies in helping users comprehend the accuracy of the model's response and calibrate their trust while aligning with their preferences. 

Our research aims to identify the most effective strategy for communicating the factuality of an LLM's response. We address the following research questions: 
 \begin{enumerate}
     \item \textbf{Trust}: Which designs foster user trust in the model?
     \item \textbf{Ease of validation}: Which designs facilitate validation of the factuality of the model's response?
     \item \textbf{Preference}: What are the most preferred designs?
 \end{enumerate}
We approached these research questions in three phases:

\begin{itemize}
    \item \textbf{Design Exploration }(Section~\ref{sec:design-exploration}): 
    We conducted a design review and a pilot study to evaluate different design options and selected a subset of those designs.
    \item \textbf{Experiment 1: Evaluative Study }(Section~\ref{sec:exp1}): We conducted a controlled scenario-based study to evaluate six design strategies for representing factuality scores. 
    \item \textbf{Experiment 2: Replication Study }(Section~\ref{sec:exp2}): We conducted a conceptual replication study~\cite{derksen2022kinds} to investigate whether the Experiment 1 findings generalize to different scenarios. % in different domains.
\end{itemize}

In the two experiments, participants were shown a color scale for conveying factuality scores, along with three styles for visualizing factuality scores within an LLM's response: (1) \textit{highlight-all}, which annotates all linguistic content in the LLM response with varying background colors based on its factuality score, (2) \textit{highlight-threshold}, which annotates only those parts of the LLM response where the factuality score is below a given threshold, and (3) \textit{score}, which shows the numeric factuality score associated with each part of the response. Factuality scores were evaluated at two levels of linguistic granularity -- \textit{phrase} and \textit{term} -- and the three factuality styles were presented at each level of granularity.

We then conducted two experiments to compare the effects of these design strategies in question-answer scenarios. We investigated their effects on participants' ratings of trust, ease of evaluating response accuracy, and preference rankings. 
In both experiments, participants had the highest preference ratings for the \textit{highlight-all} style at a \textit{phrase-level} granularity. Participants found it easier to validate the accuracy of an LLM's response in this style compared to a \textit{baseline} in which no style was applied. Moreover, displaying factuality scores led participants to increase their trust. 
Our paper makes three contributions:
\begin{enumerate}
    \item We explore the design space for presenting factuality scores to users and identify a set of promising approaches for in-depth evaluation based on user feedback.
    \item We find that design strategies significantly impact ratings of trust, ease of accuracy validation, and preference.
    \item We offer practical guidance on how to effectively communicate factuality within the user interface of LLM-based applications.
\end{enumerate}

\section{Related Work}

\subsection{LLM Hallucination and Factuality Detection}
\label{sec:relwork:factuality}
The widespread usage of LLMs in society has highlighted their risks and limitations. Notably, these models can generate text that appears plausible at first glance but actually contains factually incorrect information, a phenomenon referred to as ``hallucination" or ``confabulation.''
In contrast, \textit{factuality} is defined as ``truthfulness or the quality of being based on fact''~\cite{ji2023survey}. 
A related concept is \textit{faithfulness}, which pertains to how well an LLM-generated response is consistent with the ground truth source. In this study, we assume a reliable source as our basis for ``fact" so that faithfulness has the same meaning as factuality~\cite{maynez2020faithfulness}. 
%The source document is essential in determining the factuality of an LLM output. 
If the model's response aligns with the information from a reliable source, it is factually correct. 

Hallucinations in LLMs stem from various factors such as noisy, biased, and erroneous training data, as well as the model itself. Researchers have addressed data-related issues by establishing ground truth data through human annotators and enhancing model inputs with external knowledge~\cite{ji2023survey, huang2025survey, wang2024factuality, honovich2022true}. Efforts to enhance the model include refining the architecture (e.g., retrieval augmented generation, known as RAG;~\citealt{lewis2020retrieval}), 
improving the training process (e.g., reinforcement learning with human feedback, or RLHF;~\citealt{bai2022training}), 
and post-processing~\cite{chen2021improving}.
Each of these approaches has limitations. For instance, RAG can make statements that are not fully supported by cited sources, and may reduce the diversity of responses~\cite{liu2023evaluating}.
RLHF requires significant human labor, time, and emotional toll to refine the model~\cite{NYtimes_secret, WSJ_cleaning}. 
Given that these algorithmic approaches cannot fully ameliorate problems caused by hallucinations, in the present work, we take a human-centered perspective, emphasizing that it is the responsibility of end-users to carefully evaluate and take accountability for their use of LLM outputs.

As part of the effort to assist end-users in evaluating LLM responses, ongoing research has focused on developing methods to score the factuality of LLM outputs~\cite{laban2022summac, maynez2020faithfulness,zhou2023synthetic, chern2023factool}. These methods can either use lexical matching-based metrics relying on hard-coded logic or model-based metrics using neural networks. 
Lexical matching-based metrics, such as BLEU ~\cite{papineni2002bleu}, METEOR ~\cite{banerjee2005meteor}, and ROUGE ~\cite{lin2004rouge}, measure factuality automatically by assessing the lexical overlap between the source text and the model's response. In contrast, neural network-based metrics, including BERTscore~\cite{zhang2019bertscore}, BLEURT~\cite{sellam2020bleurt}, and FActScore~\cite{min2023factscore}, have gained popularity due to their resilience against lexical, syntactic, and semantic differences between the source and the model's output.  Moreover, task-specific model-based metrics such as ANLI \cite{nie2020adversarial}, SummaC \cite{laban2022summac}, and QuestEval \cite{scialom2021questeval} -- which are based on canonical natural language understanding tasks such as natural language inference, abstractive summarization, and question generation -- have shown promising directions for evaluating the factuality of the LLM response. 

This growing body of research raises new questions for LLM developers and designers on how to effectively \textit{communicate} factuality information to end-users. Currently, there are no established guidelines on which parts of an LLM's response should be annotated with factuality information, in what visual style, and at what level of linguistic detail. 
Furthermore, we have limited understanding of how the communication of factuality information mitigates the effects of hallucination and calibrates end-users' trust.

\subsection{Calibrating End-User Trust for AI Systems}
Successful human-AI collaboration requires a user to modulate their level of trust according to the true reliability of the AI system. This process is known as trust calibration~\cite{lee2004trust, wischnewski2023measuring, zhang2020effect}. 
Miscalibrated trust can result in overreliance, where users accept incorrect AI recommendations, or underreliance, where they reject correct recommendations from the AI.

For example, \citet{kim2023humans} asked end-users of an AI-based bird identification app about their trust and trust-related behaviors. The authors found that while people generally trusted the app, they did not accept its outputs as true every time they used the app, and carefully evaluated the outputs. If they were not able to verify the outputs due to lack of domain knowledge, participants disregarded the outputs. 
This indicates a disparity between user trust and ease of validating accuracy. In this study, we further investigate both concepts in the context of LLM interactions.

In human-human communication, uncertainty may signal transparency and honesty to a conversational partner~\cite{van2019communicating}. Similarly, in LLM interactions, communicating factuality scores or related concepts (e.g., uncertainty, confidence score) of LLM outputs can increase AI transparency~\cite{liao2023ai} and improve trust calibration~\cite{zhang2020effect}. \citet{vasconcelos2023generation} indicated that highlighting uncertain tokens can assist programmers in identifying potential errors, leading to more focused edits and greater satisfaction among study participants.  \citet{weisz2021perfection} explored a similar technique and found that confidence scores possessed explanatory power, although an analysis by \citet{agarwal2020quality} found no correlation between the model's confidence scores and the presence of actual programming errors. 
\citet{leiser2023chatgpt} found that end-users expressed a desire for visual aids such as color codes to communicate the factuality of LLM responses, which has inspired our study. 
The present work builds on this prior research and compares the effects of various design strategies for communicating the factuality of a model's response.

\section{Design Exploration}~\label{sec:design-exploration}
To develop our initial designs to present factuality scores for LLM-generated outputs, we drew inspiration from existing commercially available applications, including \mbox{OpenAI's} WebGPT %\footnote{https://openai.com/research/webgpt} 
and Microsoft's Bing Chat. %\footnote{https://www.bing.com/chat}. 
We observed that factuality information was generally presented through highlights \cite{yue2023automatic, gao2023rarr, leiser2023chatgpt} or scores \cite{li2023improving}.
In reviewing these applications, we observed that factuality information could be computed at different levels of granularity, such as at the term, phrase, or whole-response levels. 
\citet{vasconcelos2023generation} reported that users had negative reactions to the whole-response level of granularity and found it unhelpful for identifying errors and felt it was difficult to interpret. Therefore, we did not include the whole-response granularity in the present study.

We conducted a pilot study with ten participants to identify preferred options in the design space. This study led to the selection of six designs for representing factuality scores, as described in the following sections.
%shown in Table~\ref{fig:exp1-factuality_designs}. 
We then ran two controlled experiments to evaluate these different designs.

\section{Experiment 1}
\label{sec:exp1}

\subsection{Participants}\label{sec:exp1-participants}

We recruited 104 participants for this experiment, all of whom were employees of IBM, a large multinational technology company. 
Our goal was to enroll diverse participants in terms of geography, job role, English proficiency, and experience with AI, machine learning, and LLMs. We advertised the participant recruitment widely within the company on internal Slack channels from multiple divisions and geographic regions. The participants were located in 20 different countries, with the largest representation (51\%) from the United States. Their job roles encompassed various disciplines, including design, customer service, engineering, sales, research, and human resources. Participants reported a significant range of experience with LLMs, with 18.2\% indicating they had never used an LLM and 9.6\% reporting daily usage (see Figure A.1(a) for more fine-grained responses).
%(see \citealt{supplemental_mats} Figure A.1(a) for more fine-grained responses).

The experiment was conducted in English, and we strove to recruit participants with varying degrees of English exposure and proficiency to capture the experience of people who interact with LLMs in a non-native language. As such, 56\% of participants reported that they were exposed to English from birth, 27\% before age 7 (often considered the end of the critical period for learning a language to native-level proficiency; \citealt{johnson1989}), and 17\% after age 7. Participants also self-reported their English proficiency on a 7-point Likert scale, with 68\% rating themselves at \textit{7 (native or native-like proficiency)}, 19\% at \textit{6}, 8\% at \textit{5}, 4\% at \textit{4 (medium)}, and 1\% at \textit{3}.
All participants provided written informed consent, and were treated in accordance with guidelines for the ethical treatment of human participants.

% Baseline figure
\begin{figure*}[h!]
\centering
    \includegraphics[width=.75\linewidth]{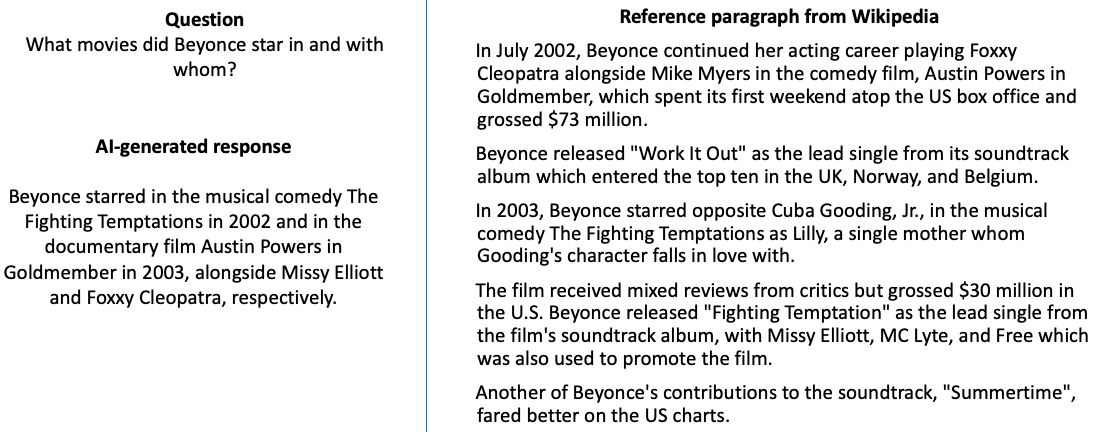}
    \caption{The \textit{baseline} design was shown to participants at the start of the experiment, with no annotations showing factuality.}
    \label{fig:baseline}
\end{figure*}

\subsection{Procedure}\label{sec:exp1-procedure}

The experimental instructions told participants to imagine themselves as users of an AI-powered language model, and were shown a Question, a Response, and a Source. Their task was to evaluate different designs for showing the factuality of the model's Response, based on the provided Source. The Question was explicitly non-technical to allow all participants, regardless of background or expertise, to assess its accuracy using the source information, and asked, ``What movies did Beyonce star in and with whom?'' The Source was an edited and condensed text pulled from the Wikipedia article about Beyonce, and participants were instructed to assume that the source was factually accurate. The Response was manually written (i.e., not actually generated by a model) to have a mixture of inaccurate statements which contradicted the source text, and accurate statements. Overall, the Response was approximately equal parts accurate and inaccurate information.

Participants were shown several design strategies to evaluate. Each design strategy was presented using the same Question, Response, and Source text, to hold constant the content and degree of accuracy across different designs. This approach allowed us to reduce the number of variables tested and ensure a more targeted exploration of the design strategies themselves. Participants always saw the \textit{baseline} design first, which had no factuality markup, and displayed the text only for the Question, Response, and Source (see Figure~\ref{fig:baseline}). 

After being shown the \textit{baseline} design, participants were asked to rate their perceptions about the model and its response on three metrics, using a 7-point Likert scale:

\begin{enumerate}
    \item \textit{Perceived accuracy}: How accurate do you think this AI-generated response is?
    \item \textit{Ease of validation}: With the information presented in this way, how easy is it for you to determine the accuracy of this AI-generated response?
    \item \textit{Trust}: With the information presented in this way, how much do you trust the AI system that generated the response?
\end{enumerate}

Following their ratings of the \textit{baseline} design, participants were introduced to the concept of a \textit{factuality score} -- a feature that compares linguistic components of the Response against the Source -- and that a high factuality score indicated that the response was aligned with the information in the Source, and thus was likely to be correct. The factuality scores in our experiment were created manually, as opposed to using an existing factuality scoring algorithm, by comparing information units in the Response against the Source text, and tweaking wording in the Response to engender a range of factuality scores to display.

\begin{figure}[b]
\centering
    %\begin{subfigure}{\columnwidth}
        \centering        
        \includegraphics[width=0.98\columnwidth]{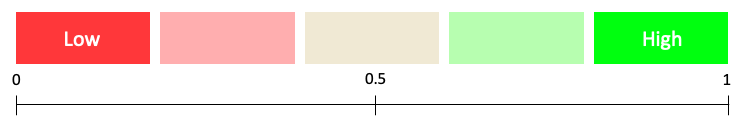}
        \label{fig:exp1-factuality_scale}
    %\end{subfigure}
\caption{The factuality scale that was presented to Experiment 1 participants. The color scale and corresponding numbers demonstrated the range of possible factuality scores.}
\label{fig:factuality_scale}
\end{figure}

Participants were then presented with six design strategies for displaying factuality information on the Response:
%Researchers hand-crafted the factuality scores of the response based on a ground truth response and the source article.
%The study was conducted as a within-subjects experiment, and thus all participants saw and rated the same designs.
%, in a randomized order to control for order effects.
Three factuality designs, each at two levels of granularity. The three designs incorporated color-coding to show the factuality of individual linguistic units in the response, on a scale ranging from 0 (red) to 1 (green)\footnote{This color scale is not ideal from an accessibility standpoint for color-blind users. We suggest modifying the color endpoints or adding shading information for systems deployed on a larger scale.}, shown in Figure~\ref{fig:factuality_scale}.
The designs were \textit{highlight-all}, in which every part of the response text was highlighted with a color corresponding to the factuality score; \textit{highlight-threshold}, in which only the sections of the response text with a factuality score below 0.5 were highlighted to signal inaccuracies; and \textit{score}, in which all parts of the response text were tagged with their factuality score, using both color-coded underlines and the numerical factuality score value.

% Factuality designs table
\begin{table*}[h]
    \centering
    \begin{tabular}{cccc}
        Granularity & Highlight-all & Highlight-threshold & Score \\
        \centeredtxt{Term} & \includegraphics[width=0.25\textwidth]{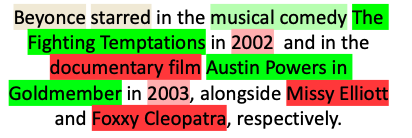} & \includegraphics[width=0.25\textwidth]{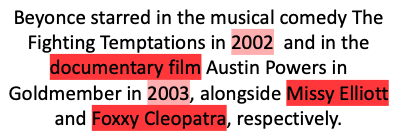} & \includegraphics[width=0.25\textwidth]{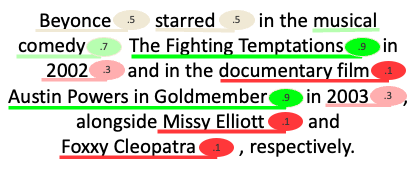} \\
        \centeredtxt{Phrase} & \includegraphics[width=0.25\textwidth]{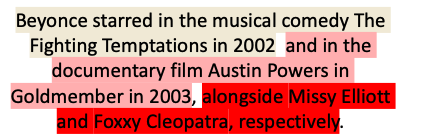} & \includegraphics[width=0.25\textwidth]{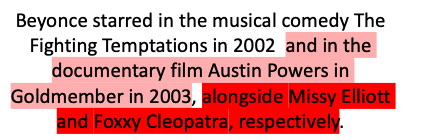} & \includegraphics[width=0.25\textwidth]{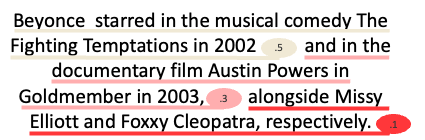} \\
    \end{tabular}
    \vspace{.3cm}
    \caption{The set of designs presented to each participant for displaying factuality scores on the model's response. Each participant saw and rated all six designs, in a randomized order grouped by granularity.}
    \label{fig:exp1-factuality_designs}
\end{table*}

In addition, designs were presented at two levels of \textit{granularity} -- \textit{term-level} or \textit{phrase-level} -- referring to the size of the text chunks over which the factuality was evaluated. At \textit{phrase-level} granularity, if there was an inaccuracy in one term in a phrase, then the entire phrase would be tagged with a lower factuality score. In contrast, at \textit{term-level} granularity, just that term would be tagged with a lower factuality score, while the other terms in the phrase or sentence would individually be tagged with their own factuality scores. Table~\ref{fig:exp1-factuality_designs} shows the six design strategies that users evaluated.

Participants saw the factuality design strategies one at a time, and rated their perceptions on two metrics: \textit{ease of validation} and \textit{trust} (questions 2 and 3 above) %in section \ref{sec:exp1-procedure}), 
on a 7-point Likert scale. Note that users were not asked about perceived accuracy (question 1) for any designs besides the \textit{baseline}, because the wording of the text was identical in all designs.

Participants performed this rating task for each of the three designs at one granularity (\textit{term-level} or \textit{phrase-level}), and then rank-ordered them along with the \textit{baseline} by preference. They then performed the same rating and preference-ranking for the three designs at the other granularity. The three designs within each granularity were presented in a randomized order across participants, and the order of the two granularities was randomized across participants to reduce order effects. %in the aggregated data.
%Following the ratings, participants were asked to explain the reasons for their preference. They were also asked a series of questions regarding interactivity of the feature -- whether they would like to have the option to selectively control when or what threshold of factuality scores are displayed.
%Following the ratings, we further asked a series of supplemental questions regarding their preference of \textit{granularity} and \textit{interactivity} of the features -- which granularity they prefer (word, phrase, others) and whether they would like to have the option to selectively control when or what threshold of factuality scores are displayed.
%Following the ratings, we asked a supplemental question regarding preference between the two types of granularity.
Finally, participants responded to demographic and professional experience questions, as reported in Section~\ref{sec:exp1-participants}.

\subsection{Analysis Methods} 
\label{exp1-analysis}

Three dependent variables that examined different facets of participants' opinions of the factuality designs were measured: 7-point Likert scale ratings of (a) trust and (b) ease of validating the response accuracy, and (c) rank-order preferences of the different designs.
Analyses used generalized linear mixed-effects models in R \cite{Rsoftware}, using the \textit{lme4}, \textit{lmerTest}, and \textit{emmeans} packages \cite{lme4, lmerTest, emmeans}, with separate models for each dependent variable.

We first assessed how each design strategy compared with the \textit{baseline} regarding ratings of trust, ease of validation, and preference. The statistical models included the within-subjects categorical independent variable Design Type, which had seven levels and was treatment-coded, with the \textit{baseline} set as the reference level. This analysis allows for the comparison of the rating of each individual design strategy against the rating given for the \textit{baseline} design. Two separate models were run, one for each of the two ratings -- trustworthiness and ease of validating accuracy -- both of which were continuous. The model's random effect structure included the levels of Design Strategy within Participant ID, with random effects which accounted for less than 1\% of the model's variance removed in order to aid convergence.
%included Participant as a random effect with Design Type nested within it to control for within-participant variance across the multiple design conditions. 
Following the full model, pairwise contrasts were conducted to explore comparisons between every pair of Design Strategy levels, with \textit{p}-values corrected for multiple comparisons using the Tukey correction.

Second, an omnibus linear mixed-effects model with the factors Granularity (term, phrase) x Design Type (highlight-all, highlight-threshold, score) was conducted, including only the six markup design strategies (i.e., excluding the \textit{baseline} design), to investigate rating differences across designs at the level of the factors Design Type and Granularity.

Finally, a cumulative link mixed model with participants' preference ranking as the dependent variable was conducted. 
Bartlett's test for homogeneity of variances indicated that the variances were not significantly different across conditions for the dependent variables. %In all models, participant ID was the random variable, and Design Strategy was the categorical independent variable.

%\RO{Update files posted on OSF:} De-identified data and scripts for running the reported analyses are available at the following repository: {\url{https://tinyurl.com/8y8e6688}}.

\subsection{Results}
\label{exp1-results}

% Exp 1 Factuality - Trust & Accuracy Figure
\begin{figure*}[h!]
\centering
    \begin{subfigure}{.95\columnwidth}
        \includegraphics[width=\linewidth]{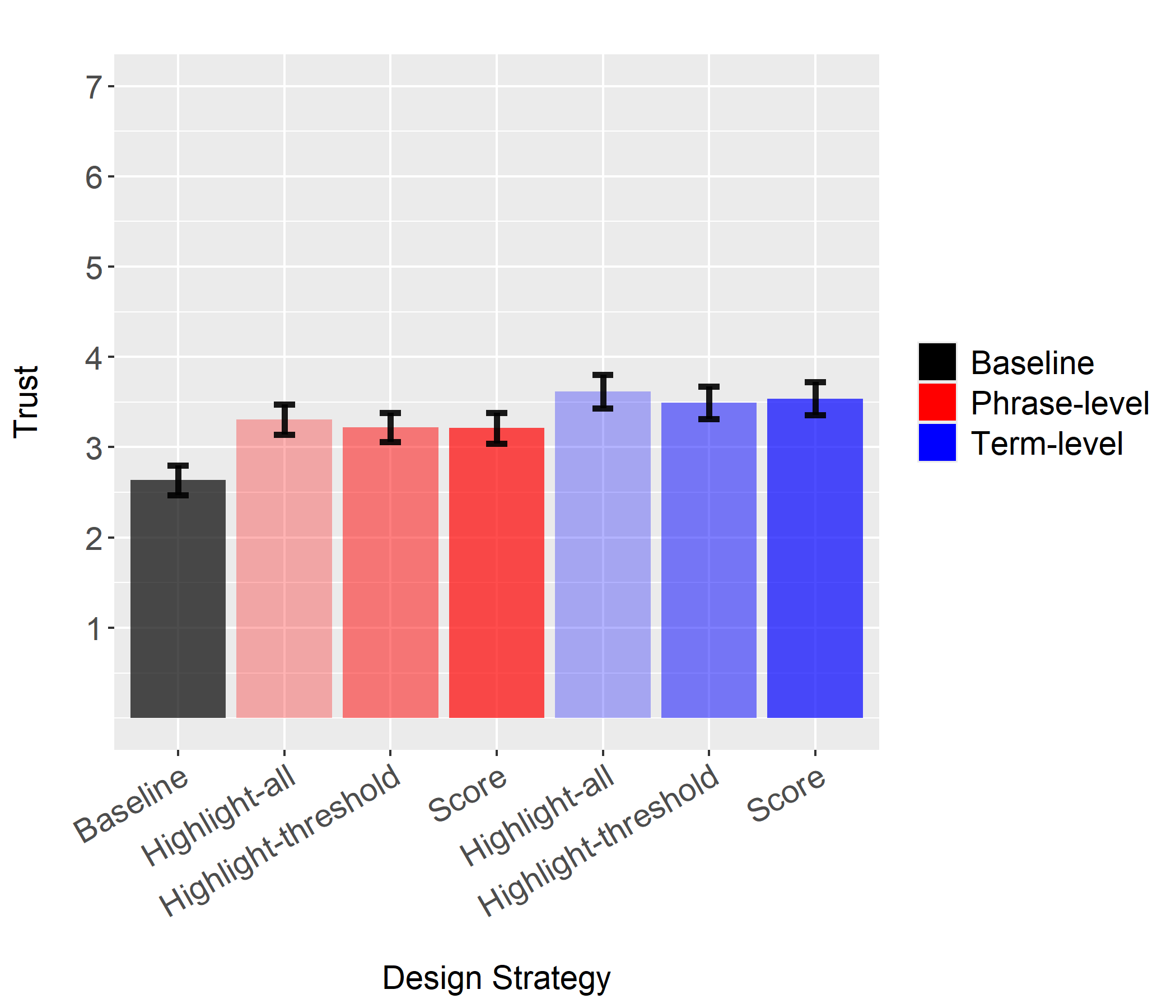}
        \caption{Trust ratings}
        \label{fig:exp1-fact_trust}
    \end{subfigure}
    \hspace{10mm}
    \begin{subfigure}{.95\columnwidth}
      \includegraphics[width=\linewidth]{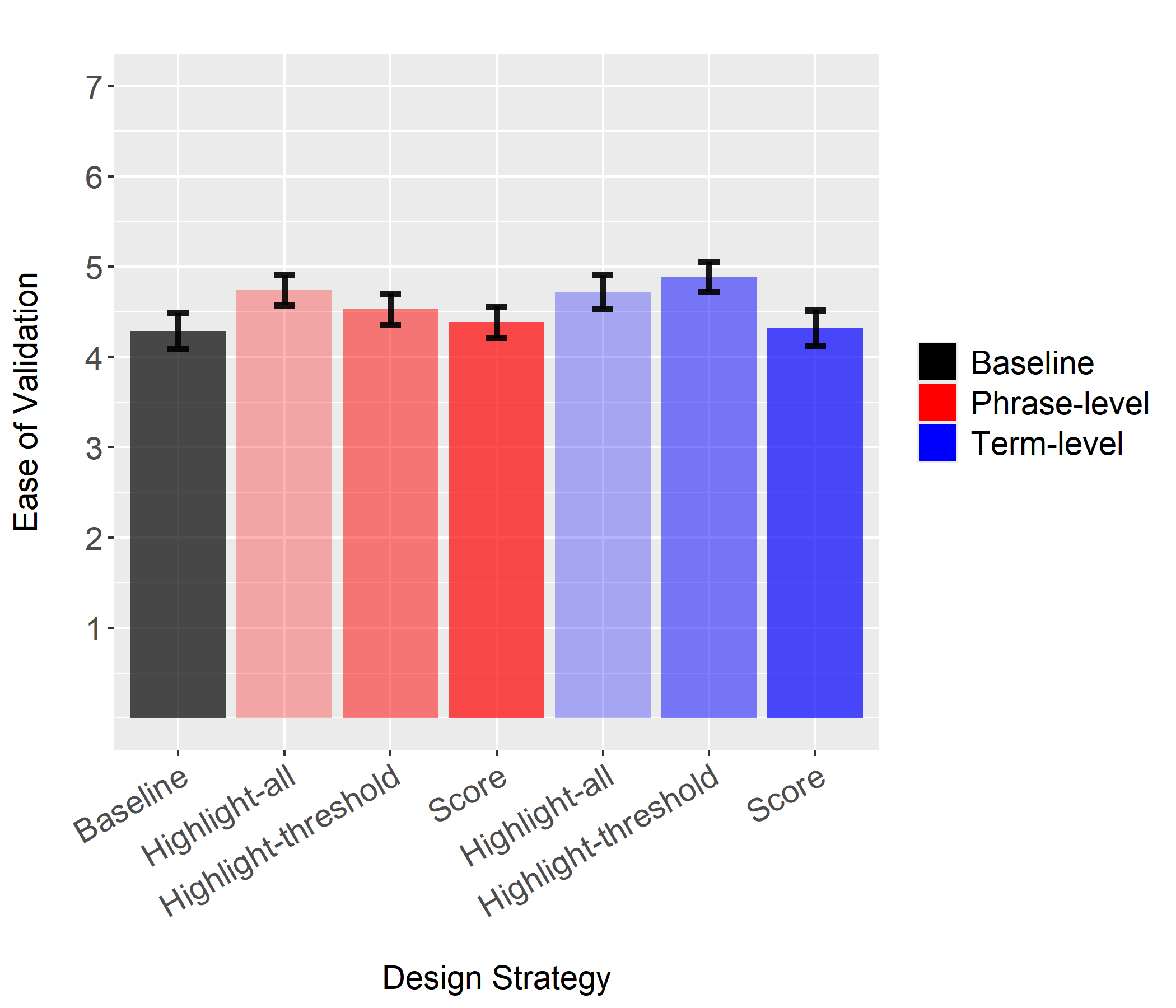}
      \caption{Ease of validating accuracy ratings}
      \label{fig:exp1-fact_accuracy}
    \end{subfigure}
\caption{Experiment 1 ratings of each design strategy: (a) trust and (b) ease of validating the accuracy of the model's response.}
\label{fig:exp1-factuality}
\end{figure*}

\subsubsection{Trust}
The first model compared participants' ratings of trustworthiness for each of the design strategies against the \textit{baseline} as the reference level. As can be seen in  Figure~\ref{fig:exp1-fact_trust}, all six designs were rated as significantly more trustworthy than the \textit{baseline}, suggesting that presenting factuality information using any of the markup methods increased participants' trust in the model. See Table~\ref{tab:ratings} for the detailed statistical results.
Post-hoc pairwise comparisons between each pair of designs revealed no additional significant differences after correction for multiple comparisons.

% Trust, Accuracy: mean, SE, t, p for both experiments
\begin{table*}[!h]
    \centering
    \scriptsize%\small
    \bgroup
    \begin{tabularx}{\linewidth}{Xp{1.4cm}p{1.65cm}p{1.4cm}p{1.65cm}p{1.4cm}p{1.65cm}p{1.4cm}p{1.75cm}}
    % \begin{tabularx}{\linewidth}{Xllll|llll}
        \toprule
         & 
        \multicolumn{4}{l}{\textbf{\textsc{Experiment 1}}} &
        \multicolumn{4}{l}{\textbf{\textsc{Experiment 2}}} \\
        \midrule 
        \textbf{Strategy} & \multicolumn{2}{l}{\textbf{Trust}} &
        \multicolumn{2}{l}{\textbf{Ease of Validation}} &
        \multicolumn{2}{l}{\textbf{Trust}} &
        \multicolumn{2}{l}{\textbf{Ease of Validation}} \\
        & \textit{Mean (SE)} & \textit{t (p)} & \textit{Mean (SE)} & \textit{t (p)} & \textit{Mean (SE)} & \textit{t (p)} & \textit{Mean (SE)} & \textit{t (p)} \\
        \midrule
        Baseline & 2.63 (0.17) & -- & 4.29 (0.20) & -- & 2.75 (0.16) & -- & 4.81 (0.17) & -- \\
        \midrule
        \textit{Phrase} \\
        highlight-all & \textbf{3.31 (0.17)} & 4.59 (\textless .001) & \textbf{4.74 (0.17)} & 2.24 (.03) & \textbf{3.67 (0.17)} & 5.81 (\textless .001) & 5.03 (0.17) & ~1.10 (.27) \\
        highlight-threshold & \textbf{3.22 (0.16}) & 4.00 (\textless .001) & 4.53 (0.17) & 1.19 (.23) & \textbf{3.38 (0.17)} & 4.15 (\textless .001) & 4.58 (0.16) & -1.15 (.25) \\
        score & \textbf{3.21 (0.17)} & 3.93 (\textless .001) & 4.38 (0.17) & 0.48 (.63) & \textbf{3.55 (0.18)} & 5.29 (\textless .001) & 4.51 (0.18) & -1.48 (.14) \\
        \midrule
        \textit{Term} \\
        highlight-all & \textbf{3.62 (0.18)} & 6.57 (\textless .001) & \textbf{4.72 (0.19)} & 2.10 (.04) & \textbf{3.34 (0.18)} & 3.84 (\textless .001) & \underline{4.23 (0.20)} & -2.77 (.01) \\
        highlight-threshold & \textbf{3.49 (0.18)} & 5.80 (\textless .001) & \textbf{4.88 (0.16)} & 2.96 (.003) & \textbf{3.18 (0.18)} & 2.87 (.004) & 4.44 (0.19) & -1.82 (.07) \\
        score & \textbf{3.54 (0.18)} & 5.85 (\textless .001) & 4.32 (0.20) & 0.13 (.89) & 2.94 (0.16) & 1.28 (.20) & \underline{3.57 (0.19)} & -6.16 (\textless .001) \\
        \bottomrule
    \end{tabularx}
    \egroup
    \caption{Trust and ease of validation ratings: means, standard errors (SE), and statistical results across the two experiments. Trust and ease of validation were rated on a 1-7 Likert scale, with 7 as the highest score. \textit{t}- and \textit{p}-values are from the model with the \textit{baseline} set as the reference level.
    %and thus indicate whether each factuality design was rated significantly differently than the \textit{baseline}. 
    %Bolded black text indicates ratings that were significantly higher than the \textit{baseline}, and bolded blue text indicates ratings that are significantly less than the \textit{baseline}.
    Bolded text indicates ratings that were significantly higher than the \textit{baseline}, and underlined text indicates ratings that were significantly lower than the \textit{baseline}.
    }
    \label{tab:ratings}
\end{table*}

Next, we conducted the omnibus model of Granularity x Design Type. This revealed a main effect of Granularity (\textit{F(1,~103)}~=~6.30, \textit{p}~=~.01), with \textit{term-level} granularity designs (M~=~3.54) rated higher than \textit{phrase-level} designs (M~=~3.25). There was no main effect of Design Type (\textit{F(2,~106)}~=~1.27) or interaction (\textit{F(2,~309)}~\textless~1).

\begin{figure}[h!]
\centering
    \includegraphics[width=\linewidth]{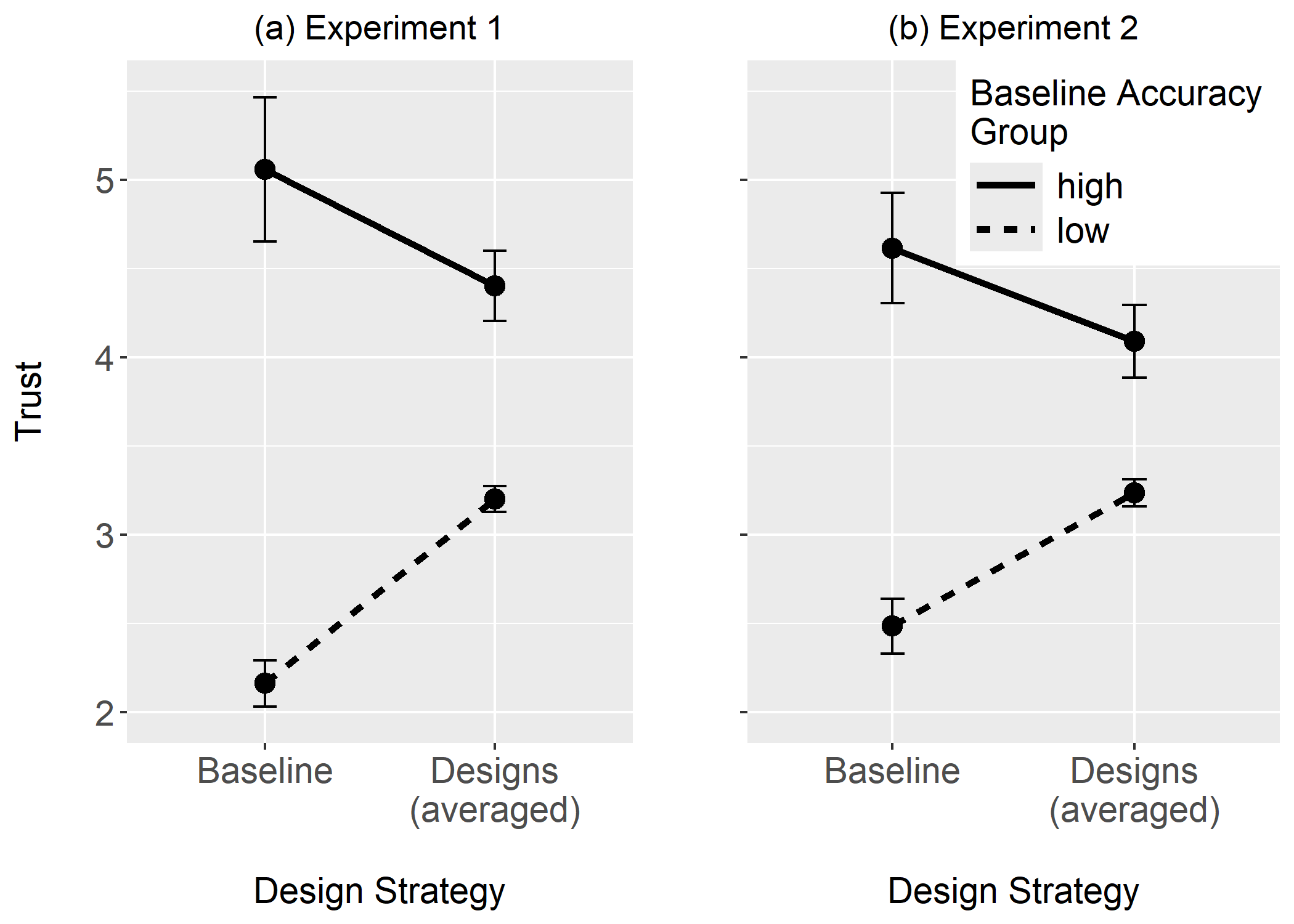}
    \caption{Participants' trust ratings for the \textit{baseline} design, and for the six design markups averaged together, for (a) Experiment 1 and (b) Experiment 2. Participants were divided into two groups based on whether they rated the model's response as high accuracy (rating of 5-7) or low accuracy (rating of 1-4) when initially presented with \textit{baseline} design.}
    \label{fig:baseline_acc_split}
\end{figure}

As an exploratory post-hoc analysis, we investigated whether participants' rating of the accuracy of the model's response when presented with the \textit{baseline} design (question (1) in Section~\ref{sec:exp1-procedure}) affected how much they trusted the model when it subsequently displayed factuality scores. 
For visualization purposes, we categorized participants into two groups: 
The \textit{low baseline accuracy} group, which was defined as those participants who rated the model's response accuracy in the \textit{baseline} design at or below 4 (the midpoint of the rating scale; N~=~87), and the \textit{high baseline accuracy} group, who rated the \textit{baseline} response accuracy as 5 or higher (N~=~17). %Note that the variable used to categorize participants into the two groups (accuracy of the model's response) is different than the variable plotted in this figure (trust of the model which produced the response).
As can be seen in Figure~\ref{fig:baseline_acc_split}(a), participants who rated the model's response at \textit{baseline} with low accuracy (dashed line) also had low trust of the model when viewing the baseline non-marked-up design, but subsequently increased their trust ratings after reviewing the factuality score markups. 
In contrast, participants who initially felt the model's response was more accurate (solid line) also had high trust of the model's response when shown the \textit{baseline} design, but subsequently \textit{decreased} their trust ratings after examining the factuality information. % presented in each design style.
%Taking one step further, we explored the relationship between the perceived accuracy of the model's response in baseline and participants' self-rated frequency of using LLMs. After excluding the data from one participant who reported their frequency as `don't know,' we built a linear regression model. The frequency of LLM usage predicted participants' perceived accuracy of the baseline with marginal significance (b = 0.15, t(101) = 1.74, \textit{p}~<~.1).

% Split-by-baseline accuracy (trust calibration) figure - Exp 1 & 2

\subsubsection{Ease of validation} 

The first model compared participants' ratings on the ease of assessing the model's accuracy for each of the designs against the \textit{baseline}. As can be seen in Figure~\ref{fig:exp1-fact_accuracy}, of the six design strategies, three were rated as significantly easier to assess the response accuracy compared to the \textit{baseline}: \textit{highlight-all} at phrase-level granularity, and \textit{highlight-all} and \textit{highlight-threshold} at term-level granularity. The other three designs were not significantly different from the \textit{baseline}; see Table~\ref{tab:ratings} for the detailed statistical results. Post-hoc pairwise comparisons between each pair of design strategies revealed no significant differences after correction for multiple comparisons. %The mean and standard errors of the ratings for each design, along with results from the statistical model comparing each design to the \textit{baseline}, are displayed in Figure~\ref{fig:exp1-fact_accuracy} and Table~\ref{table:exp1-factuality-accuracy}. 

Next, the omnibus model investigating Granularity x Design Type revealed a main effect of Design Type (\textit{F(2,~103)}~=~4.62, \textit{p}~=~.01), with \textit{highlight-all} (M~=~4.73) and \textit{highlight-threshold} (M~=~4.71) rated higher than \textit{score} (M~=~4.35).
There was a Granularity x Design Type interaction (\textit{F(2, 206)}~=~3.12, \textit{p}~=~.05), the locus of which can be seen in Figure~\ref{fig:exp1-fact_accuracy}, largely driven by the fact that \textit{highlight-threshold} was rated higher relative to the other designs at the \textit{term-level} but not \textit{phrase-level}. 
%; see Table~\ref{table:exp1-factuality-accuracy}). 
There was no main effect of Granularity (\textit{F(1, 103)}~\textless~1).

\subsubsection{Preference}
Participants rank-ordered each of the three designs and the \textit{baseline} within each granularity level. Thus, each ranking response compared four designs, with rank 1 for the most preferred and 4 for the least preferred design. %The data here are presented with the ranking scores reversed (i.e., M = 4 - ranking score) such that a higher score corresponds to a more preferred design, for easier comparability with the accuracy and trust ratings in the previous analyses. %At phrase-level granularity, the \textit{highlight-all} design was the most preferred (M = 2.06, SE~=0.10), \textit{score} was second most preferred (M = 2.31, SE~=~0.10), \textit{highlight-threshold} was third (M = 2.33, SE = 0.08) and the baseline was the least preferred (M = 3.31, SE = 0.12). 
At \textit{phrase-level} granularity, the \textit{highlight-all} design was the most preferred, \textit{score} was second most preferred, \textit{highlight-threshold} was third, and the \textit{baseline} was the least preferred (see Table~\ref{tab:preference_rankings}).
There was a significant effect of Design Type on the means of preference rankings (${\chi}^2$(3)~=~103.82, \textit{p}~\textless~.001). 
Post-hoc pairwise comparisons revealed that participants preferred each of the factuality designs significantly more than the \textit{baseline} (\textit{p}~\textless.001). 
Similarly, at \textit{term-level} granularity, the \textit{highlight-all} design was the most preferred, \textit{highlight-threshold} was second, \textit{score} was third, and \textit{baseline} was the least preferred.
There was a significant effect of Design Type on the means of the preference rankings (${\chi}^2$(3)~=~78.01, \textit{p}~\textless.001). 
Post-hoc pairwise comparisons revealed that participants preferred each of the factuality designs significantly more than the \textit{baseline} (\textit{p}~\textless.001). 
No other comparisons yielded statistically significant results.

Participants were also asked about their preference between the two types of granularities: 52.9\% of participants preferred \textit{phrase-level} granularity, while 26.9\% of the participants preferred \textit{term-level} granularity, with 10.6\% of participants responding with ``don't know'' and 9.6\% of participants selecting ``other'' (e.g. sentence-level, entire response).

% Preference ranking: mean, SE for both experiments
\begin{table*}[h!]
    \centering
    \scriptsize%\small
%\resizebox{\columnwidth}{!}{
    \begin{tabular}{ l|cc|lcc }
        \toprule
        & \multicolumn{2}{c|}{\textbf{\textsc{Experiment 1}}} &
        \multicolumn{3}{c}{\textbf{\textsc{Experiment 2}}} \\
        %Design Strategy & Mean &  SE & \textit{t} \\
        \textbf{Strategy} & \textbf{Phrase} &  \textbf{Term} & \textbf{Phrase} &  \textbf{Term} & \textbf{Overall} \\
        \midrule
        Baseline  & 3.31 (0.12) & 3.22 (0.12)  & 3.34 (0.10)& 2.90 (0.13)& 5.14 (0.21)  \\
        \midrule
        \multicolumn{1}{l|}{\textit{Phrase} } & \multicolumn{2}{l|}{}\\
            Highlight-all        & \textbf{2.06 (0.10)}  & -- & \textbf{1.82 (0.09)}& -- & \textbf{2.56 (0.15)}   \\
            Highlight-threshold  & 2.33 (0.08)  & -- & 2.51 (0.09)& -- &3.62 (0.17)  \\
            Score                & 2.31 (0.10) & -- & 2.34 (0.10) & -- & 3.56 (0.19)   \\
        \midrule
       \multicolumn{1}{l|}{\textit{Term} } & \multicolumn{2}{l|}{}\\
            Highlight-all        & -- & \textbf{2.13 (0.10)}  & -- & \textbf{2.09 (0.09) }& 3.94 (0.16) \\
            Highlight-threshold  & -- & 2.21 (0.09) & -- & 2.16 (0.09) &4.02 (0.19)  \\
            Score                & -- & 2.44 (0.10) & -- &2.85 (0.10)& 5.16 (0.17) \\
        \bottomrule
    \end{tabular}%}
    \caption{Preference rank-order means (standard errors). The most preferred design is bolded.
    Ranking scores are on an inverse scale; thus lower numbers indicate higher preference.
    In both experiments, participants ranked designs separately within each granularity level. 
    In Experiment 2, participants additionally ranked designs across both granularity levels in a single question.}
    \label{tab:preference_rankings}
\end{table*}

\section{Experiment 2} \label{sec:exp2}

The goal of Experiment 2 was to assess whether the Experiment 1 results were robust to extension, and to automatize aspects of generating the Response and factuality scores. As such, several aspects of the procedure were modified. 
First, Experiment~1 investigated factuality designs in a single question-answer scenario; a goal of Experiment~2 was to assess the generalizability of the results to additional domains. Thus Experiment 2 included two new scenarios: a medical question and an HR question. These were selected as domains for which LLMs are increasingly employed for business automation purposes. 
%The two new scenarios are shown in Figure~B.1 in \citet{supplemental_mats} Appendix B.
The two new scenarios are shown in Figure~B.1 in Appendix B.

Second, in contrast to Experiment 1 where the Response was manually generated without LLM input, 
for Experiment 2, we started each scenario with an LLM-generated response as ground-truth (after comparing it to the Source and determining that it was entirely faithful).
As this experiment required the presented Response to have a range of factuality scores, we then edited the model's ground-truth response to contain errors, which became the Response for the experimental scenario. To determine the factuality scores to display in the designs, we used the Python spaCy library~\cite{spacy} 
to calculate the semantic similarity between each term or phrase in the edited Response and the corresponding term or phrase in the ground-truth response, and used that similarity as the factuality score for that linguistic unit. Thus the factuality scores shown in Experiment 2 were
generated by automated means rather than human-created as in Experiment 1. 
We opted to create the Response and factuality scores in this manner (rather than using an entirely-live LLM response and factuality scoring algorithm) as it gave us experimental control over the range of factuality scores that was presented to participants, as the goal of the present experiment was to assess opinions about the factuality \textit{designs}, rather than assessing the accuracy of factuality scoring \textit{algorithms}.

These factuality scores were then used to generate the markups for the six design strategies. To determine the highlight colors, factuality scores were mapped to colors in a modified manner from that in Experiment 1. As the semantic similarity-based factuality scores were biased towards high-magnitude, positive values, 
the color mapping thresholds were adjusted to have higher resolution at the parts of the scale with the largest concentration of numerical scores.
%See \citet{supplemental_mats}~Figure~B.2 
See Figure~B.2 
for the modified Experiment~2 factuality color scale.

Minor adjustments were made to the linguistic units used for factuality scores: In Experiment 1, we annotated some multi-word noun phrases (e.g., ``musical comedy'' or ``documentary film'') as one term, whereas for Experiment 2, term-level markups were only individual words with the exception of multi-word proper nouns (e.g., disease names). 
We also added a question asking participants to rank all six designs together, in addition to ranking the designs separately within each level of granularity.

\subsection{Participants}
\label{exp2-participants}
We recruited another 104 IBM employees via internal Slack channels. A condition of participation was that they had not participated in Experiment 1. 
Participants' work locations consisted of 17 unique countries, with the US as the most common (59\%). Job roles again spanned a wide array of disciplines, and participants had a  range of experience with LLMs, from never to daily usage
(see Figure~A.1(b)).
%(see \citealt{supplemental_mats} Figure~A.1(b)).
Participants had varying degrees of English exposure and proficiency, with a very similar distribution as in Experiment 1: 58\% of participants reported that they were exposed to English from birth, 19\% before age 7, and 23\% after age 7. For self-rated proficiency, 71\% rated themselves at \textit{7 (native or native-like proficiency)}, 20\% at \textit{6}, 5\% at \textit{5}, 3\% at \textit{4 (medium)}, and 1\%~at~\textit{3}.
All participants provided written informed consent and were treated in accordance with the guidelines for ethical treatment of human participants.

\subsection{Procedure}
The procedure for Experiment 2 was largely the same as that of Experiment 1, with a few changes as noted above. 
%To further explore participant preference of granularity, we introduced an additional question that solicited a comprehensive ranking of \textit{all} designs across both the term-level and the phrase-level, rather than their preference within each level of granularity. \todo{JW: it wasn't an additional question, it was a tweak to the previous ranking questions.. we should make that clearer. Maybe we rewrite this to, ``We adjusted how we asked participants to rate their preferred designs by having them rate all six designs at once, rather than having them separately rate the designs within each level of granularity. This approach allowed us to make more direct comparisons between each highlighting scheme across the granularities.''}
%The analysis methods were similar to those in Experiment 1. %except that we added the scenario type (medical vs. HR) as a covariate in all models.
Participants were randomly assigned to one of the two scenarios, HR or medical, and rated all of the designs for that scenario 
(see Figure~B.1 for the text of the scenarios). 
%(see \citealt{supplemental_mats} Figure~B.1 for the text of the scenarios). 
There were 55 participants who completed the HR scenario, and 49 participants who completed the medical scenario. %Due to some participants beginning the experiment but not completing it, assignment to the two scenarios was not exactly balanced; only data from participants who completed the full experiment were analyzed.

\subsection{Results} \label{exp2-results}

\subsubsection{Trust}

\begin{figure*}[h!]
 \centering
    % \begin{subfigure}{.6\columnwidth}
     \begin{subfigure}{.4\textwidth}
         \includegraphics[width=\linewidth]{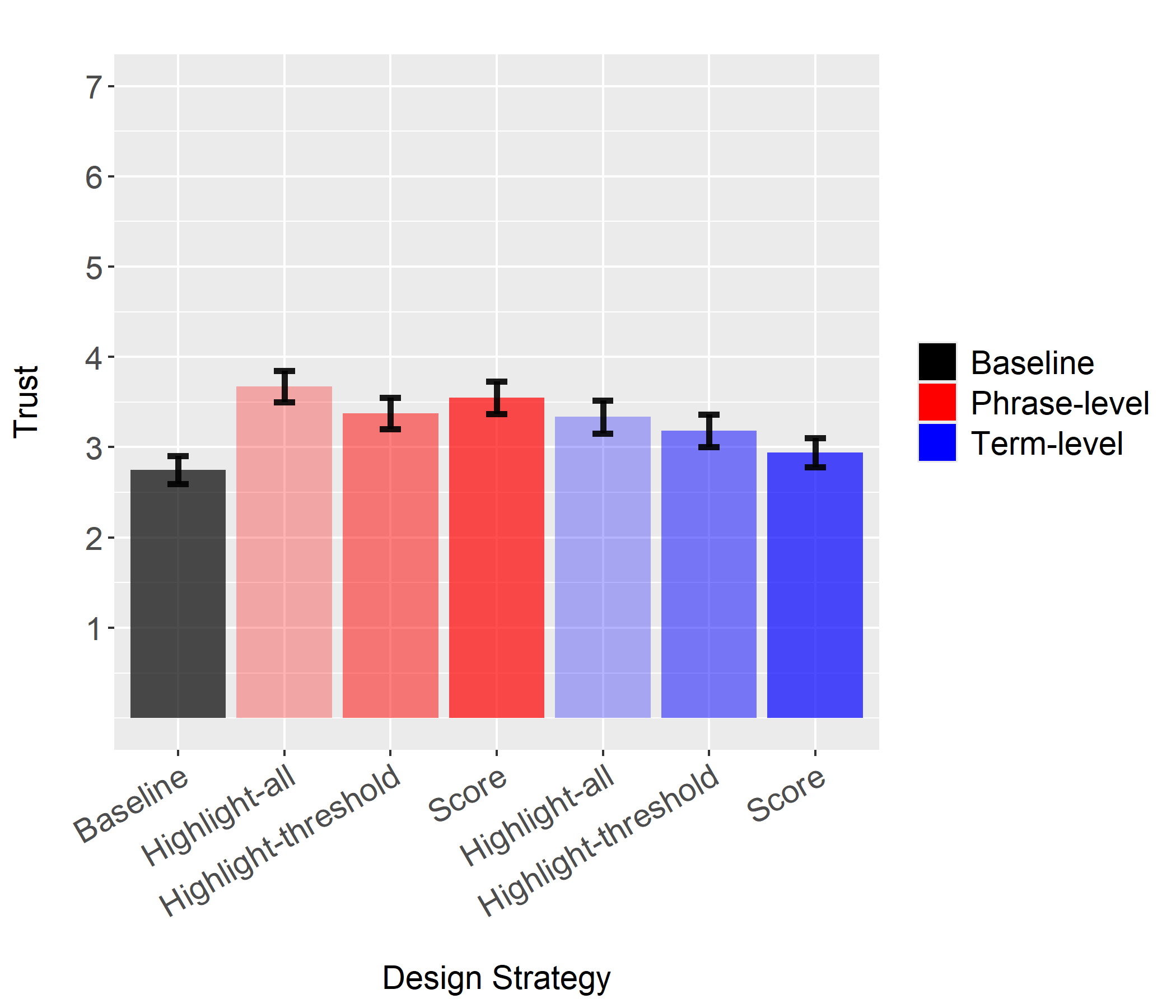}
         \caption{Trust ratings}
         \label{fig:exp2-fact_trust}
     \end{subfigure}
     \hspace{10mm}
   % \begin{subfigure}{.6\columnwidth}
     \begin{subfigure}{.4\textwidth}
       \includegraphics[width=\linewidth]{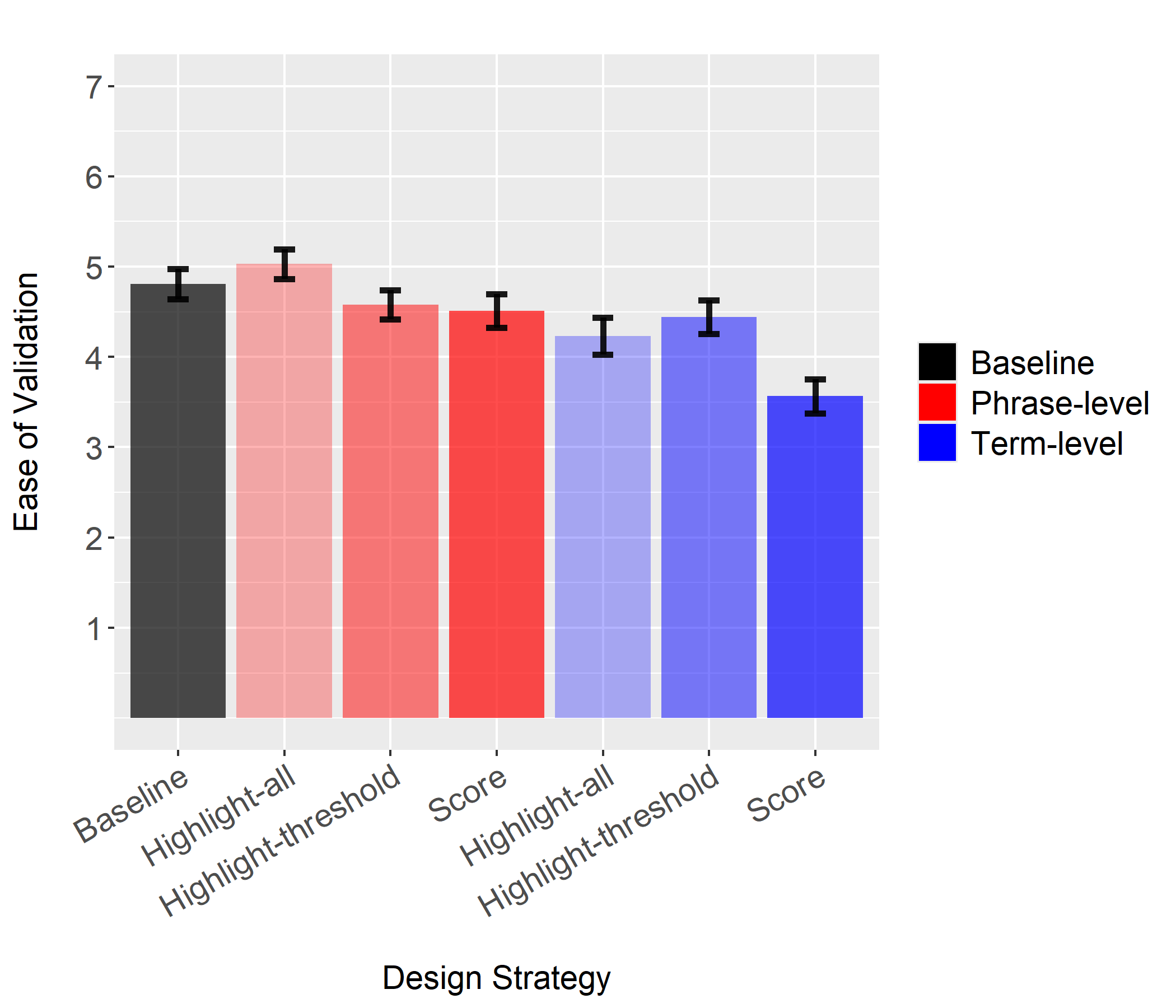}
       \caption{Ease of validating accuracy ratings}
       \label{fig:exp2-fact_accuracy}
     \end{subfigure}
 \caption{Experiment 2 ratings of each design strategy: (a) trust and (b) ease of validating the accuracy of the model's response.}
 \label{fig:exp2-factuality}
 \end{figure*}

First, we ran a model comparing the trustworthiness ratings of each of the design strategies against the \textit{baseline}. All of the designs with the exception of \textit{term-level score}
%All three of the \textit{phrase-level} designs, as well as \textit{highlight-all} and \textit{highlight-threshold} at the \textit{term-level} 
were rated significantly higher on trust than was the \textit{baseline} design (see Figure~\ref{fig:exp2-fact_trust}). The mean and standard error of the trust ratings, as well as results from this statistical model comparing each design strategy against the \textit{baseline}, are shown in Table~\ref{tab:ratings}. In the post-hoc pairwise comparisons between all pairs of designs, three remained significant after correcting for multiple comparisons: \textit{phrase-level highlight-all} was rated as significantly more trustworthy than both \textit{term-level highlight-threshold} (\textit{t(240)}~=~3.09, \textit{p}~=~.04) and \textit{term-level score} (\textit{t(240)}~=~4.60, $\textit{p}~<.001$), and \textit{phrase-level score} was rated as significantly more trustworthy than \textit{term-level score} (\textit{t(437)}~=~4.02, \textit{p}~=~.001).

 % \section{Experiment 2 Results}~\label{appendix:exp2_results}
 % \setcounter{figure}{0}

Next, we conducted an omnibus model investigating the main effects of Granularity and Design Type on the six design strategies, excluding the \textit{baseline}. There was a main effect of Granularity (\textit{F(1, 103)}~=~7.73, \textit{p}~=~.007), with \textit{phrase-level} granularity (M~=~3.53) rated higher than \textit{term-level} granularity (M~=~3.15).
There was a main effect of Design Type (\textit{F(2, 125)}~=~6.24, \textit{p}~=~.003), with \textit{highlight-all} rated the highest (M~=~3.50), and \textit{highlight-threshold} (M~=~3.28) and \textit{score} (M~=~3.25) rated lower.
There was also a Granularity x Design Type interaction (\textit{F(2, 243)}~=~4.08, \textit{p}~=~.02). The locus of this interaction can be seen in Figure~\ref{fig:exp2-fact_trust} and Table~\ref{tab:ratings}, by the difference in ratings of the \textit{score} design strategy between \textit{phrase-level} and \textit{term-level} Granularity.

To investigate whether participants found the different scenarios (i.e., HR vs. medical scenario) to affect their trust ratings, we additionally ran a Granularity x Design Type x Scenario model on the ratings of the six design strategies (excluding the \textit{baseline}). As with the previous model, there was a main effect of Granularity (\textit{F(1, 103)}~=~8.18, \textit{p}~=~.005) and Design Type (\textit{F(2, 140)}~=~6.24, \textit{p}~=~.003), as well as a Granularity x Design Type interaction (\textit{F(2, 231)}~=~3.93, \textit{p}~=~.02). However, importantly, all effects involving the Scenario factor were not significant (all \textit{p}s~\textgreater.12), demonstrating that idiosyncracies of the scenarios themselves were not driving the trust ratings.

We again conducted an exploratory visualization, splitting participants into two groups based on how they had rated the accuracy of the model's response when presented with the initial \textit{baseline} design: high accuracy (N~=~13) or low accuracy (N~=~91); see Figure~\ref{fig:baseline_acc_split}(b). As in Experiment 1, participants who thought the model's response had low accuracy (dashed line) also initially rated the model with low trust, but subsequently increased their trust ratings when presented with design information detailing the factuality of the response. In contrast, participants who thought the model's response had high accuracy (solid line) initially rated the \textit{baseline} design with high trust, but decreased their trust ratings once presented with factuality score information.

\subsubsection{Ease of validation}
First, we ran a model comparing the rating for each of the six design strategies against the \textit{baseline} rating. Although two design strategies were rated significantly different than the \textit{baseline} design -- \textit{term-level highlight-all} and \textit{term-level score} -- both were rated lower, that is, as more difficult to validate the accuracy of compared to the \textit{baseline}. See Figure~\ref{fig:exp2-fact_accuracy} and Table~\ref{tab:ratings} for the numerical and statistical results. 
In the post-hoc pairwise comparisons between all pairs of designs, all of the design strategies, including the \textit{baseline}, were rated as easier to validate the accuracy than \textit{term-level score} (all \textit{t}s~\textgreater~3.18, all \textit{p}s~\textless~.03). In addition, \textit{phrase-level highlight-all} was rated significantly higher than \textit{term-level highlight-all} (\textit{t(259)}~=~3.83, \textit{p}~=~.003).

The omnibus model of Granularity x Design Type for the six design strategies (excluding the \textit{baseline}) revealed a main effect of Granularity (\textit{F(1, 103)}~=~12.83, \textit{p}~\textless.001), with \textit{phrase-level} granularity (M~=~4.71) rated higher than \textit{term-level} granularity (M~=~4.08). There was a main effect of Design Type (\textit{F(2, 124)}~=~15.03, \textit{p}~\textless~.001), with \textit{highlight-all} rated the highest (M~=~4.63), \textit{highlight-threshold} (M~=~4.51) in the middle, and \textit{score} (M~=~4.04) rated lowest.
There was also a Granularity x Design Type interaction (\textit{F(2, 243)}~=~9.78, \textit{p}~\textless~.001), as can be seen in the different patterns of Design Strategy in Figure~\ref{fig:exp2-fact_accuracy} and Table~\ref{tab:ratings}.

Next, we ran a Granularity x Design Type x Scenario model to investigate whether the different stimulus scenarios affected ease of validation ratings. The results were very similar to the previous model, with a main effect of Granularity (\textit{F(1, 102)}~=~12.72, \textit{p}~\textless~.001), a main effect of Design Type (\textit{F(2, 123)}~=~14.94, \textit{p}~\textless~.001), and a Granularity x Design Type interaction (\textit{F(2, 240)}~=~9.83, \textit{p}~\textless~.001). There was a main effect of Scenario (\textit{F(1, 102)}~=~5.84, \textit{p}~=~.02), with the Medical scenario (M~=~4.71) rated easier to validate than the HR scenario (M~=~4.07). This likely reflects the fact that the medical scenario was noticeably shorter than the HR scenario, so there was less text to read through and verify the accuracy. %(see Figure~\ref{fig:exp2-baseline}). 
However, no interactions involving Scenario reached significance, demonstrating that properties of the scenarios themselves did not differentially affect ratings of validation ease between the different design strategies, and thus the conclusions of the relative ratings of the design strategies hold across multiple scenarios.

\subsubsection{Preference}
Preferences, ranked from most (1) to least (7) preferred across both granularity levels, were as follows: \textit{phrase-level highlight-all}, \textit{phrase-level score}, \textit{phrase-level threshold}, \textit{term-level threshold}, \textit{term-level highlight-all}, \textit{baseline}, and \textit{term-level score}.
There was a significant effect of Design Strategy on the means of preference rankings (${\chi}^2$(6)~=~150.42, \textit{p}~\textless.001). Post-hoc pairwise comparisons showed that participants significantly preferred the \textit{phrase-level highlight-all} design over all other designs (\textit{p}~\textless.01). In contrast, both \textit{baseline} and \textit{term-level score} designs were significantly less preferred compared to other designs, and there was no significant difference between the two. See Table~\ref{tab:preference_rankings} for the ranking means.

We additionally aggregated the preference rankings by Granularity level.
There was a significant effect of Granularity on the means of preference rankings (${\chi}^2$(2)~=~100.88, \textit{p}~\textless.001). The preference rankings followed the order of \textit{phrase-level}, \textit{term-level}, and \textit{baseline}, with every pairwise comparison showing a significant difference (\textit{p}~\textless.001).

\section{Discussion}

In the present work, we created six design strategies for displaying factuality about an LLM's response in a question-answer scenario, and conducted two experiments
where participants rated these designs on trust, ease of validating the accuracy, and preference. Overall, highlighting every phrase in the response using a factuality score color scale (the \textit{phrase-level highlight-all} design) was the most preferred, trusted, and easiest for users to validate the accuracy of a response.
Our results suggest several design recommendations for communicating factuality scores to users, which we explain according to each outcome, and also 
additional factors that may influence factuality designs.

\subsection{Design Recommendations}

\textbf{Trust} All of our factuality designs were effective at increasing and calibrating trust compared to the baseline of showing no factuality information. 
%One exception was the term-level \textit{score} design, for which we did not find a significant difference compared to the \textit{baseline} in Experiment 2. One possible reason is that the scores cluttered the response more in Experiment 2. There were more score annotations in Experiment 2 responses than in the Experiment 1 response, because the responses were longer, and the definition of ``term'' was stricter. One participant in Experiment 2 said, ``\textit{The numbers in colored bubbles makes the text difficult to read}.'' This implies that it is essential to balance transparency with clarity.
%While prior explainable AI research has mixed findings about the relationship between explanations and trust, our findings support that explanations can help increase user trust. 
%Procedural justice theory~\cite{lind1988social} suggests that people's trust is strongly impacted by procedural explanations and not only the outcome. Our design strategies provided a procedural explanation of how the model's response was generated, even where the response was partially inaccurate.
Hence, we recommend presenting factuality information using one or more of the proposed designs to increase user trust. 
In addition, in an exploratory analysis, participants' initial accuracy assessment of the model's response %in the \textit{baseline} design
had a substantial impact on their trust after they viewed the factuality scores.
Participants who initially overlooked errors in the model's response \textit{decreased} their trust after viewing the errors called out through the factuality scores. In contrast, participants who initially identified errors in the model's response \textit{increased} their trust when they observed that the factuality scores accurately flagged those errors. 
%This pattern of trust calibration based on their initial assessment of response accuracy echoes findings from prior research that individuals whose expectations were violated trusted the system less~\cite{kizilcec2016much}. 
%Expectancy violations theory supports the finding that positive violations (i.e., when perceived performance exceeds rather than meets the expected level of performance) have a stronger positive effect on satisfaction, while negative violations produce a negative effect~\cite{burgoon2016application}. 
%We also observed preliminary findings suggesting a marginally positive relationship between individuals' accuracy assessments of the \textit{baseline} and their frequency of using LLMs. Participants who have had successful interactions using LLMs are more likely to become frequent users, and this positive bias might lead them to interpret the model's output as more accurate than it actually is.
Therefore, the current results suggest that incorporating factuality information into LLM responses might help to appropriately calibrate the level of end-users' trust in the model -- either in a positive or a negative direction.
%again support incorporating factuality information in the LLM response. 
%LLM designers and developers may also consider adaptive interfaces in response to expectation violations to induce high trust, which is likely influenced by the frequency of LLM usage. We encourage further research exploration in this direction.
%Therefore, to induce high trust, these results suggest that LLM designers and developers should consider adaptive interfaces in response to expectation violation, providing factuality scores to users with low expectations. One potential approach is to provide factuality scores to newcomers who are likely to have less experience using the LLM and have low expectations. We encourage further research exploration in this direction.
%One potential approach involves providing factuality scores to users with more exposure to AI, who may have lower expectations. We encourage further research exploration in this direction.

\textbf{Ease of validation} 
The \textit{phrase-level highlight-all} design was rated as easier to validate the model's accuracy than the \textit{baseline} design. %across the two experiments.
However, designs at \textit{term-level} granularity showed inconsistent results between the two experiments: In Experiment 1, \textit{term-level highlight-all} and \textit{highlight-threshold} designs were rated as significantly easier to assess accuracy than the \textit{baseline}. In contrast, in Experiment 2, the \textit{term-level highlight-all} and \textit{score} designs were rated as more difficult to validate accuracy than the \textit{baseline}. 
A possible explanation is that the number of annotated terms was higher in Experiment 2 (HR: 39, Medical: 17) than in Experiment 1 (10), potentially overwhelming participants' ability or interest in investigating the accuracy of the model's response.
A participant in Experiment 2 mentioned, ``\textit{The numbers in colored bubbles makes the text difficult to read}.'' 
Additionally, the proportion of inaccurate terms (color-coded with red or pink) was lower in Experiment 2 (HR: 33\%, Medical: 35\%) than in Experiment 1 (50\%),
%Since only inaccurate terms are useful for cross-checking with the source during validation, the lower proportion of such terms in Experiment 2 
which may have led users to perceive that \textit{term-level} designs are less informative for validating accuracy. 
As a result, we recommend adjusting the granularity of annotations according to the model's response length, or incorporating a feature that enables users to disable or filter accurate terms, in order to reduce distractions when assisting users in validating the accuracy of the response.

%\citet{kim2023humans} found that while people generally trusted their AI-powered app, they did not accept its outputs as true every time they used the app and carefully evaluated the outputs. If they were not able to verify the outputs, participants disregarded the outputs, despite their positive assessment of the app's trustworthiness. This indicates a disparity between user trust ratings and their trust-related behavior -- in this case, the validation of output accuracy. Along with this prior work, our data substantiates the assertion that our designs did not help verify response accuracy, despite being perceived as trustworthy. Therefore, we recommend incorporating a feature that enables users to turn off or filter styles to reduce distractions, or even remove the styles if prioritizing ease of validation over building trust.

\textbf{Preference} %When asked about their preferences regarding the granularity of presenting the factuality information within a model's response, 
The most common granularity preference %regarding the granularity of presenting factuality information 
was \textit{phrase-level}, and within the phrase-level granularity designs, \textit{highlight-all} was most preferred. Thus, we recommend the \textit{highlight-all} style in cases where user preference and experience is the primary objective.

\subsection{Factors Impacting Factuality Communication}
Our study investigated scenarios where factuality information is likely to be valued; namely, a question-answer task. However, LLM ``hallucinations'' are not always undesirable or factually inaccurate. For example, researchers have explored the use of LLMs as creativity support tools for tasks including writing stories~\cite{wang2023popblends}, fan fictions~\cite{alfassi2025online}, and humor~\cite{wu2025one}; brainstorming and ideation~\cite{muller2024group, muller2024workplace, he2024ai}; and using analogical thinking to solve design problems~\cite{yang2025overload}. In use cases like these, presenting factuality scores are likely less useful, because the model's response is not expected to be an accurate or veridical reflection of a source (or reality).
%may not be useful because hallucinations might potentially foster creativity. 

While our study assumed that the algorithm generating factuality scores for the model's response is reliable, no algorithm is perfect and incorrect factuality scores have the potential to erode users' trust in a model. 
One way to mitigate this issue is to present factuality scores in a fuzzy format, such as using ranges, instead of precise numbers. In our study, we found that the highlighting methods that visually represent a range of factuality scores with a single color performed better than the score method, which presented the precise numerical factuality score. It is possible that this fuzziness was one factor in why the highlighting designs were generally favored over the score designs.
Therefore, we encourage HCI researchers to further investigate how to effectively communicate the uncertainty in factuality scores to address the limitations of the algorithms.
%the algorithm itself may be imperfect. Incorrect factuality scores have the potential to erode users' trust in a model. 
%Therefore, it is important to ensure the reliability of the factuality score algorithm before adopting it in LLM interfaces. %\TODO{JW: this places too much burden on ML communities. HCI communities can instead look at how to communicate the uncertainty in factuality scores since we know they won't be perfect.}

The present research focused on a model's response which contained inaccuracies, but it is important to acknowledge situations where a response is entirely accurate or faithful to the source. % and when a question is unanswerable from the information in the source document~\cite{yue2023automatic}. 
High factuality scores should increase end-users' confidence in the LLM response, although some users may be skeptical when seeing a perfect rating. 
%In the case of an unanswerable question -- when the source document does not provide information that answers the user's question -- \TBD{WHAT?}
% to determine correctness, our current design strategies are not applicable since there is no response to annotate or attribute. 
Additionally, in the present work, we told participants to assume the source was reliable; however, it is possible for a response to be \textit{faithful} to an unreliable source. In such cases, high factuality (or faithfulness) scores may be misleading and could lead to over-reliance on the response. We encourage future research to explore design strategies for various situations.

\subsection{Limitations and Future Directions}
Our experiments focused on a question-answer scenario. An important avenue for future research is to explore factuality designs in other LLM tasks, such as summarization or classification.
% \TBD{varied response characteristics: more domains, variable length, etc}
While Experiment 2 tested whether our conclusions generalized to two additional scenarios, there is an enormous amount of unexplored variation in 
%While we tested additional scenarios in the second experiment, no study can capture the full range of variations of 
domains and LLM interactions in real-world scenarios. Future research should explore additional topic domains, source and response lengths, and other features of the interaction. %and structures. 
%Like any other method, the survey method and within-subject experimental design have trade-offs, such as a limited capacity to measure behavioral indicators and interactions. We encourage researchers to explore other research methods to enhance the reliability of the findings.

%We also handcrafted the model's response, rather than generating an actual LLM response.
We also created the experimental Responses by editing a real LLM response rather than using the original response.
This allowed us to create designs with varying factuality scores that effectively tested our research questions. However, real-world LLM responses may be different, such as emitting a one-word response. 
However, considering the rapid pace of technological change, the responses may vary over time and across models, making it less critical to rely on actual responses produced by existing models. 

While we made efforts to recruit participants with diverse skills, LLM experience, language proficiency, and geographic locations, our participants were all employees of a single technology company.
Future studies should involve broader participant samples from the general public. 

Finally, it is important to note that our research did not aim to exhaustively explore all potential design strategies. Instead, this study should be viewed as a starting point, encouraging researchers to delve deeper into diverse design strategies and expand the discussion.

\section{Conclusion}

Large language models have known problems with hallucinations. To address these challenges, researchers are developing algorithms to assess the factuality of an LLM model's output, but how to effectively communicate such factuality information to end-users %to help them calibrate their trust 
is an open question. We conducted two experiments using three different scenarios to compare six design strategies for communicating factuality scores against a no-markup baseline. % and assessed their impact on trust, ease of validation, and user preference.
% Through two scenario-based experiments, in which we varied scenarios and factuality scoring methods,
We found consistent results showing that \textbf{highlighting every phrase in the model's response} based on its factuality score was the most preferred strategy, %than the baseline without any markups. 
and led to high trust of the model.
This design was also perceived to be easier to validate the accuracy of the LLM response than the baseline. 
Our findings also suggested that factuality designs may enable participants to appropriately calibrate their trust in a model. %after their initial accuracy assessment of the response. 
%We recommend that future research further investigate different types of tasks, presentations of factuality scores, and variations in the accuracy of responses and sources.
%Based on our findings, we recommend our factuality designs, particularly the \textit{phease-level highlight-all} approach to enhance trust and align with user preferences.
Thus, presenting factuality score information in an understandable way is an important tool for end-users to be able to evaluate properties of large language models that are critical to being an informed consumer of AI.

% \section*{Acknowledgements}

%% If your work has an appendix, this is the place to put it.
\newpage
\onecolumn
\appendix

\renewcommand\thefigure{\thesection.\arabic{figure}}

\section{Participant Experience}~\label{appendix:demogs}

\setcounter{figure}{0}

 \begin{figure*}[h]
 \centering
     \begin{subfigure}{.4\textwidth}
         \includegraphics[width=\linewidth]{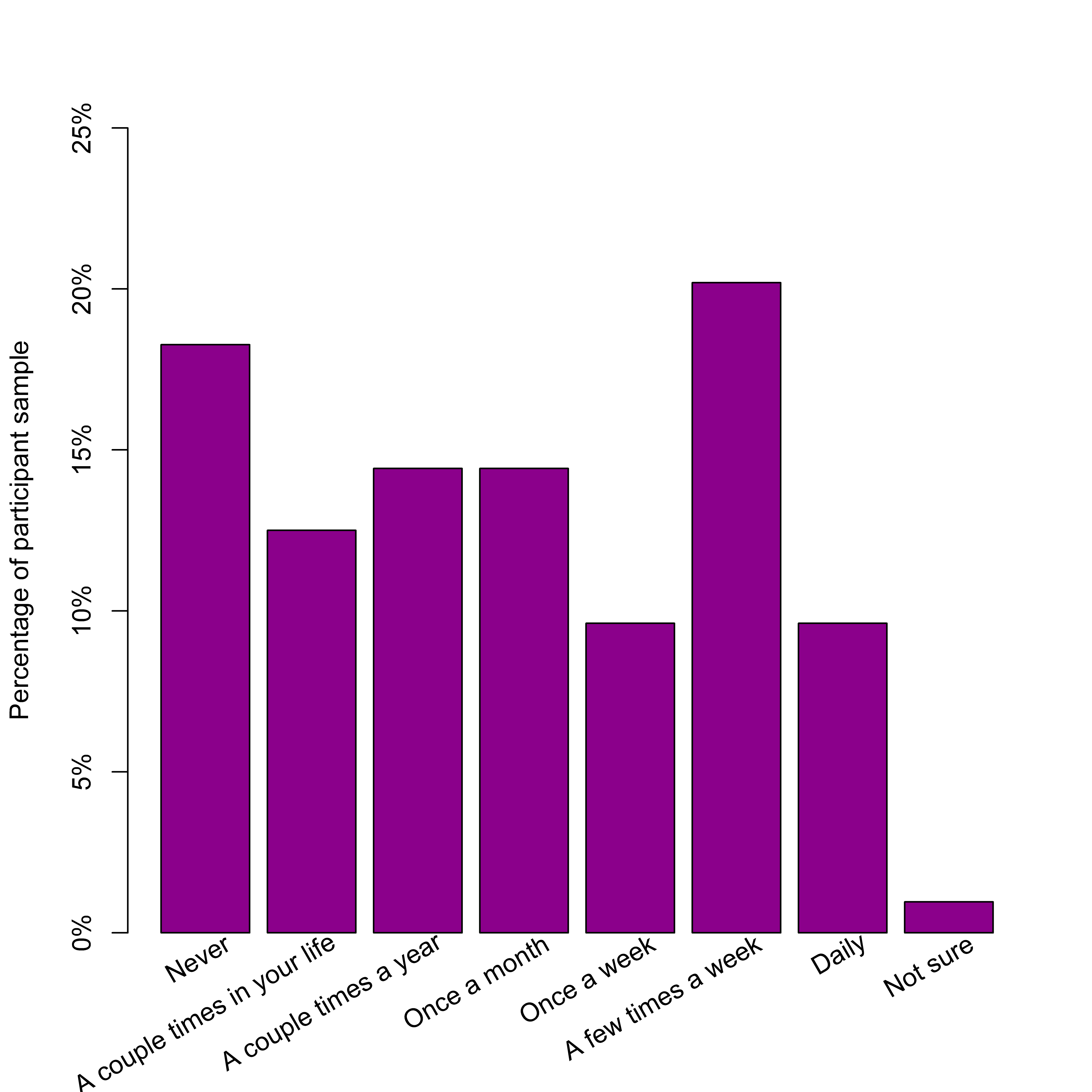}
         \caption{Experiment 1 participants}
         \label{fig:Exp1-LLM_experience}
     \end{subfigure}
     \hspace{10mm}
     \begin{subfigure}{.4\textwidth}
       \includegraphics[width=\linewidth]{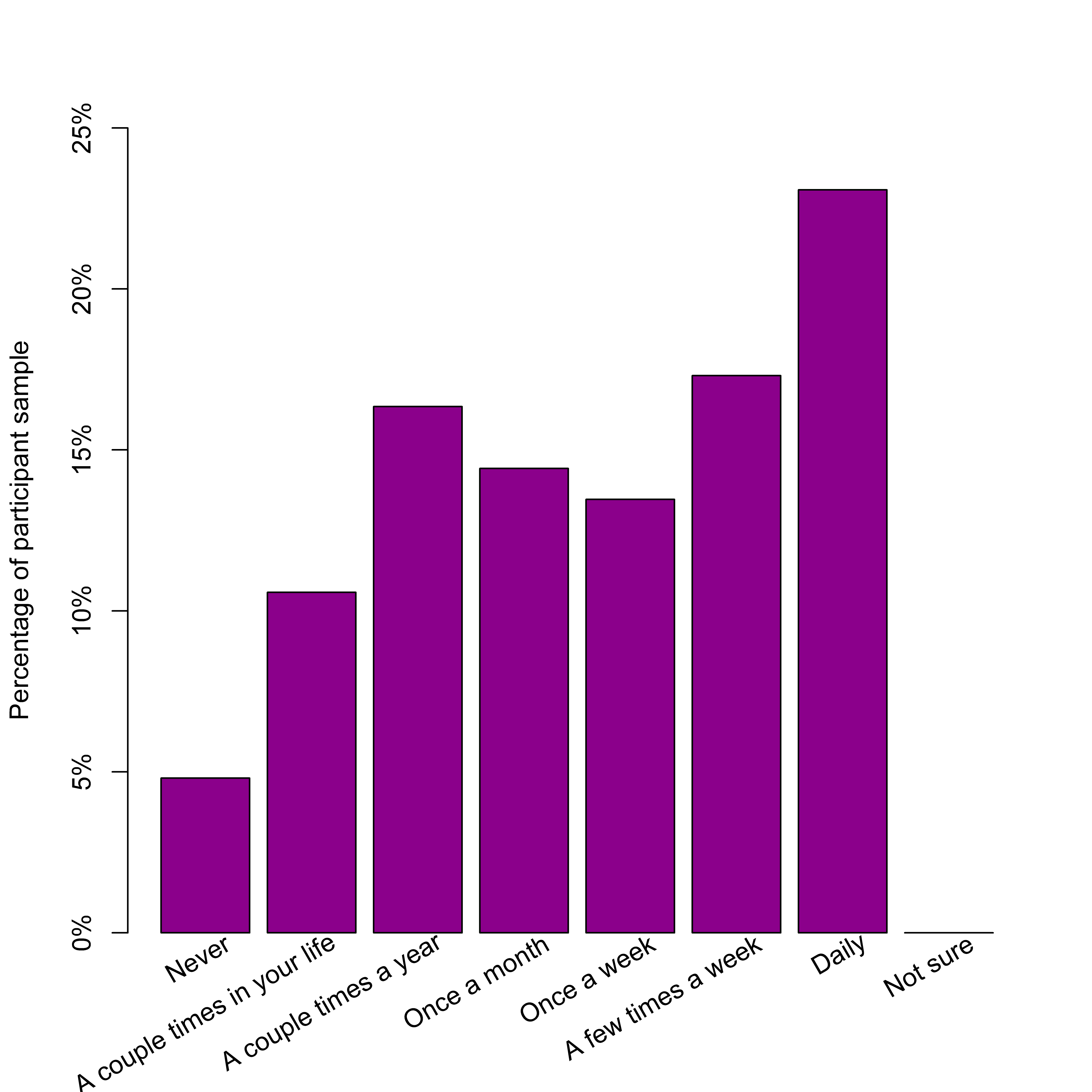}
       \caption{Experiment 2 participants}
       \label{fig:Exp2-LLM_experience}
     \end{subfigure}
 \caption{Participants' response to the question \textit{How often do you use large language models like ChatGPT, Bard, and Bing Chat? Either as part of your studies, your job, or as a hobby.}}
 \label{fig:LLM_experience}
 \end{figure*}

\newpage
\section{Experiment 2 Materials}~\label{appendix:exp2}
\setcounter{figure}{0}

%\noindent\begin{minipage}{\textwidth}
 % Exp 2 baseline designs
 \begin{figure*}[h]
 \centering
  \begin{subfigure}{.63\linewidth}
     \centering
      \includegraphics[width=\linewidth]{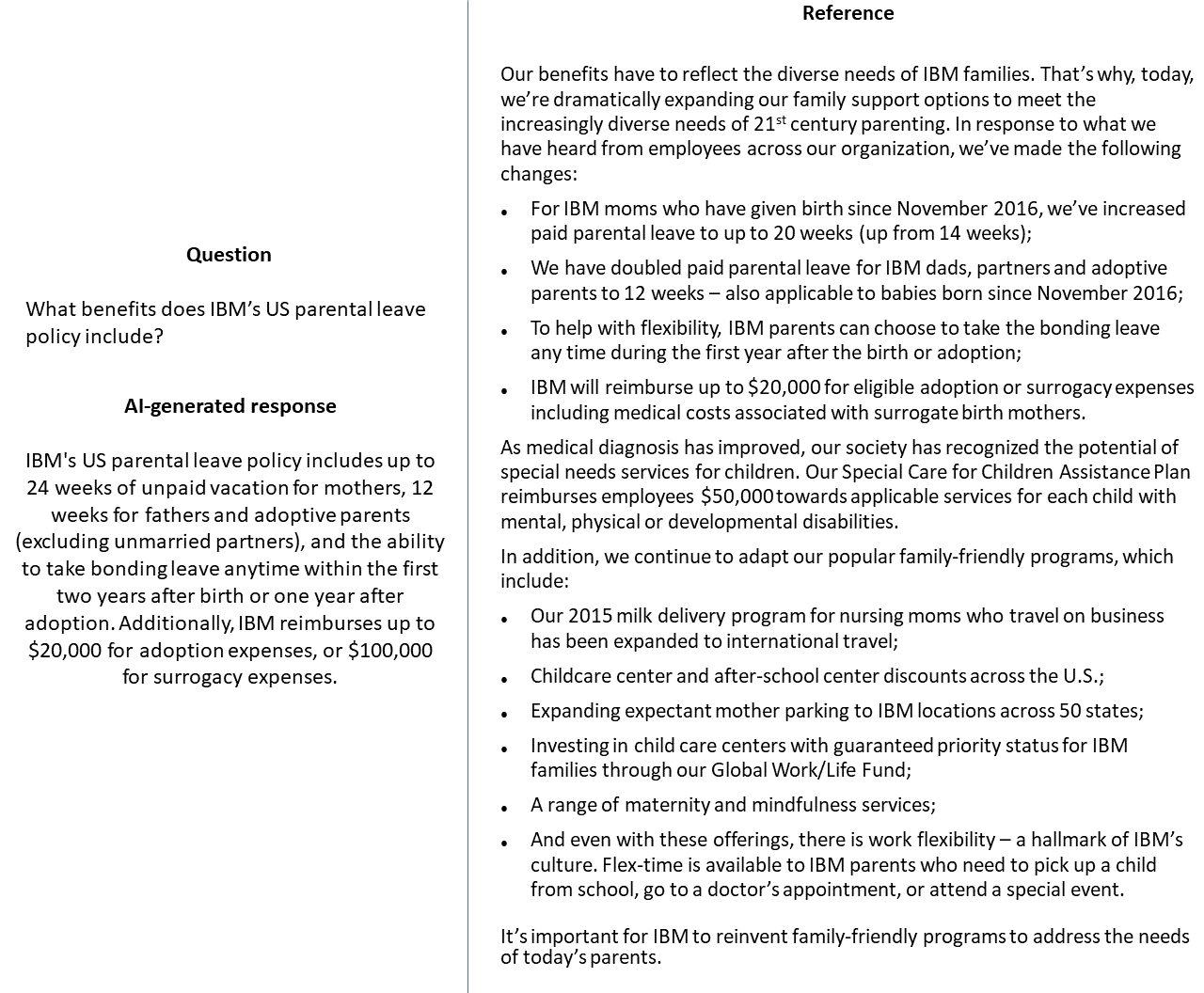}
      \caption{}
      %\label{}
      %\vspace{.2cm}
  \end{subfigure}
  \begin{subfigure}{.63\linewidth}
     \centering
    \includegraphics[width=\linewidth]{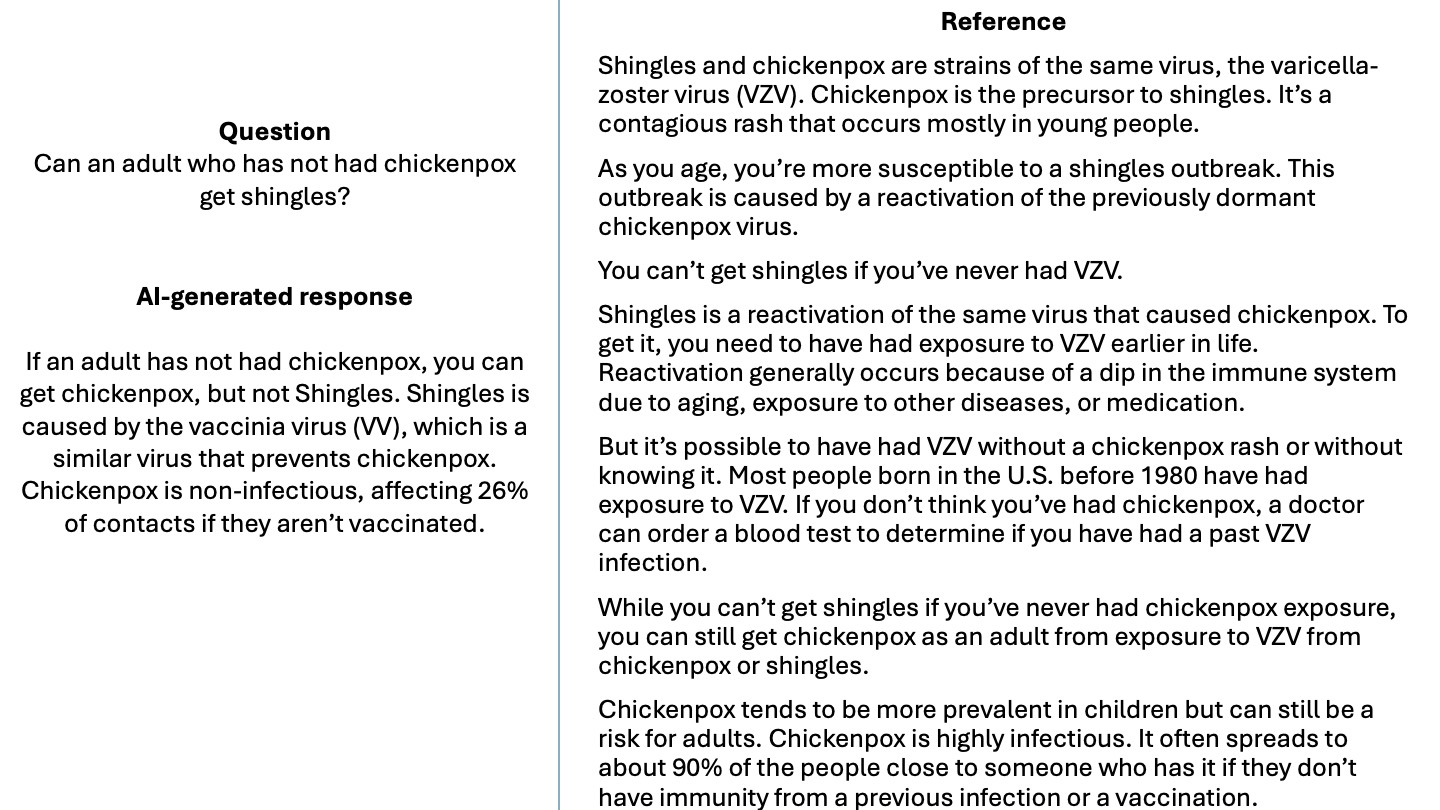}
    \caption{}
  \end{subfigure}
 \caption{The \textit{baseline} design for the two scenarios in Experiment 2. (a) The HR scenario. (b) The medical scenario text was drawn from an LLM-generated response in~\citet{kim2024m}.}
 \label{fig:exp2-baseline}
 \end{figure*}

%\noindent\begin{minipage}{\textwidth}
\begin{figure*}[t]
\centering
    %\vspace{.5cm}
    \includegraphics[width=0.7\linewidth]{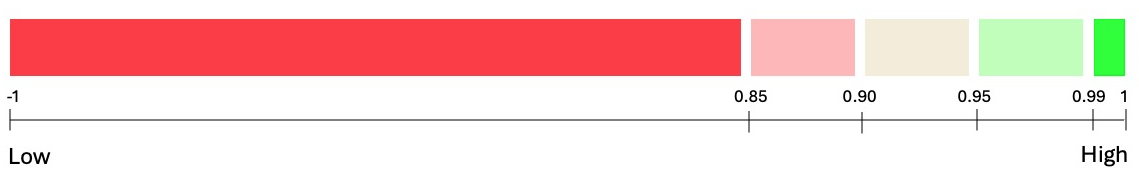}
    \caption{Factuality scale used in Experiment 2. The color thresholds were adjusted based on the distribution of factuality scores.}
    \vspace{128in}
    \label{fig:Exp2-factuality_scale}
\end{figure*}
%\end{minipage}

%%
%%
%% The next two lines define the bibliography style to be used, and
%% the bibliography file.
% \bibliographystyle{aaai25}
\twocolumn
\bibliography{references}

\end{document}